\documentclass[sigconf]{acmart}
\usepackage{subfigure}
\usepackage{multirow}
\AtBeginDocument{%
  \providecommand\BibTeX{{%
    \normalfont B\kern-0.5em{\scshape i\kern-0.25em b}\kern-0.8em\TeX}}}

\copyrightyear{2021}
\acmYear{2021}
\setcopyright{acmcopyright}\acmConference[MM '21]{Proceedings of the 29th ACM International Conference on Multimedia}{October 20--24, 2021}{Virtual Event, China.}
\acmBooktitle{Proceedings of the 29th ACM International Conference on Multimedia (MM '21), October 20--24, 2021, Virtual Event, China}
\acmPrice{15.00}
\acmDOI{10.1145/3474085.3475605}
\acmISBN{978-1-4503-8651-7/21/10}

\begin{document}
\fancyhead{}
%%
%% The "title" command has an optional parameter,
%% allowing the author to define a "short title" to be used in page headers.
\title{Two-pronged Strategy: Lightweight Augmented Graph Network Hashing for Scalable Image Retrieval}

%%
%% The "author" command and its associated commands are used to define
%% the authors and their affiliations.
%% Of note is the shared affiliation of the first two authors, and the
%% "authornote" and "authornotemark" commands
%% used to denote shared contribution to the research.
%\author{Hui Cui$^{1}$, Lei Zhu$^{1}$, Jingjing Li$^{2}$,  Zhiyong Cheng$^{3}$, Zheng Zhang$^{4}$}
% \affiliation{
% \institution{$^{1}$ Shandong Normal Unversity, Jinan, China}
% \institution{$^{2}$ University of Electronic Science and Technology of China, Chengdu, China}
%   \country{China}
% \institution{$^{3}$ Shandong Academy of Sciences, China}
% \institution{$^{4}$ Harbin Institute of Technology, Shenzhen, China}
% }
%\email{leizhu0608@gmail.com}
%\thanks{Lei Zhu is the corresponding author.}

\author{Hui Cui}
 \affiliation{
 \institution{Shandong Normal University}
  \city{Jinan}
  \country{China}}

\author{Lei Zhu$^{*}$}
 \affiliation{
 \institution{Shandong Normal University}
  \city{Jinan}
  \country{China}}
\thanks{Lei Zhu (leizhu0608@gmail.com) is the corresponding author.}

\author{Jingjing Li}
\affiliation{%
  \institution{University of Electronic Science and Technology of China}
   \city{Chengdu}
   \country{China}}

\author{Zhiyong Cheng}
\affiliation{%
  \institution{Shandong Artificial Intelligence Institute}
   \city{Jinan}
   \country{China}}

\author{Zheng Zhang}
\affiliation{%
 \institution{Harbin Institute of Technology}
  \city{Shenzhen}
  \country{China}}

%%
%% By default, the full list of authors will be used in the page
%% headers. Often, this list is too long, and will overlap
%% other information printed in the page headers. This command allows
%% the author to define a more concise list
%% of authors' names for this purpose.
\renewcommand{\shortauthors}{Trovato and Tobin, et al.}

%%
%% The abstract is a short summary of the work to be presented in the
%% article.
\begin{abstract}
Hashing learns compact binary codes to store and retrieve massive data efficiently. Particularly, unsupervised deep hashing is supported by powerful deep neural networks and has the desirable advantage of label independence. It is a promising technique for scalable image retrieval. However, deep models introduce a large number of parameters, which is hard to optimize due to the lack of explicit semantic labels and brings considerable training cost. As a result, the retrieval accuracy and training efficiency of  existing unsupervised deep hashing are still limited. To tackle the problems, in this paper, we propose a simple and efficient \emph{Lightweight Augmented Graph Network Hashing} (LAGNH) method with a two-pronged strategy. For one thing, we extract the inner structure of the image as the auxiliary semantics to enhance the semantic supervision of the unsupervised hash learning process. For another, we design a lightweight network structure with the assistance of the auxiliary semantics, which greatly reduces the number of network parameters that needs to be optimized and thus greatly accelerates the training process. Specifically, we design a cross-modal attention module based on the auxiliary semantic information to adaptively mitigate the adverse effects in the deep image features. Besides, the hash codes are learned by multi-layer message passing within an adversarial regularized graph convolutional network. Simultaneously, the semantic representation capability of hash codes is further enhanced by reconstructing the similarity graph. Experimental results show that our method achieves significant performance improvement compared with the state-of-the-art unsupervised deep hashing methods in terms of both retrieval accuracy and efficiency. Notably, on MS-COCO dataset, our method achieves more than 10\% improvement on retrieval precision and 2.7x speedup on training time compared with the second best result.
\end{abstract}

%%
%% The code below is generated by the tool at http://dl.acm.org/ccs.cfm.
%% Please copy and paste the code instead of the example below.
%%

%%
%% Keywords. The author(s) should pick words that accurately describe
%% the work being presented. Separate the keywords with commas.

%% A "teaser" image appears between the author and affiliation
%% information and the body of the document, and typically spans the
%% page.
%%
%% This command processes the author and affiliation and title
%% information and builds the first part of the formatted document.
\maketitle

\section{Introduction}
Hashing \cite{survey2016,survey2018,zhu3,zhu5} aims to project the high-dimensional data into low-dimensional and similarity-preserving binary codes, which can greatly reduce the consumption of data storage space and accelerate the retrieval process. Due to its desirable advantages, hashing has received extensive attentions recently, and various methods for large-scale image retrieval have been proposed.

Pioneering hashing methods are based on shallow learning models. They generally achieve sub-optimal performance, as they separate the feature representation and hash code learning into two independent processes \cite{AGH2011,ITQ2013,SGH2015,LSMH2017,KSH2012,LFH2014,SDH2015,FastH2015}. With the success of deep learning and representation learning \cite{add2,add3,add4,zhu1,zhu2,zhu4}, recent hashing methods have moved towards deep hashing which simultaneously learns feature representations and hash functions within a unified deep neural network. Supervised deep hashing methods \cite{CNNH2014,NINH2015,DPSH2016,DCH2018} train the deep hashing models on labeled data and preserve the semantics from explicit labels into the hash codes. They rely on a large amount of labeled data, which impedes the scalability of supervised hashing. In contrast, unsupervised deep hashing methods \cite{UH-BDNN2016,DH2017unsuper,GreedyHash2018,DistillHash2019} learn hash functions and hash codes without this reliance, and thus they support scalable image retrieval well.

Initially, unsupervised deep hashing methods usually enforce several important criteria on the hash layer of the deep neural network to learn the hash functions \cite{DeepBit2016,DH2017unsuper}. Since they fail to preserve the semantic similarities between images into the hash codes during the hash learning process, their performance is far from satisfactory. To address this problem, several approaches use image reconstruction strategy as an indirect way to preserve the similarities of images, such as BGAN \cite{BGAN2018} and TBH \cite{TBH2020}. Besides, several approaches are proposed to construct the semantic similarity structures based on the image features as the self-supervised information to guide the hash model training, such as SSDH \cite{SSDH2018}, SADH \cite{SADH2018}, and DU3H \cite{DU3H2020}. Nevertheless, unsupervised deep hashing is very difficult to optimize and it suffers from low training efficiency, as a large number of parameters is involved in the deep neural network and no explicit semantic supervision can be provided. How to reduce the number of parameters to be optimized and simultaneously enhance the semantic representation capability of deep networks are two important problems that need urgent solutions.
\begin{figure*}
\centering
\subfigure{\includegraphics[scale=0.45]{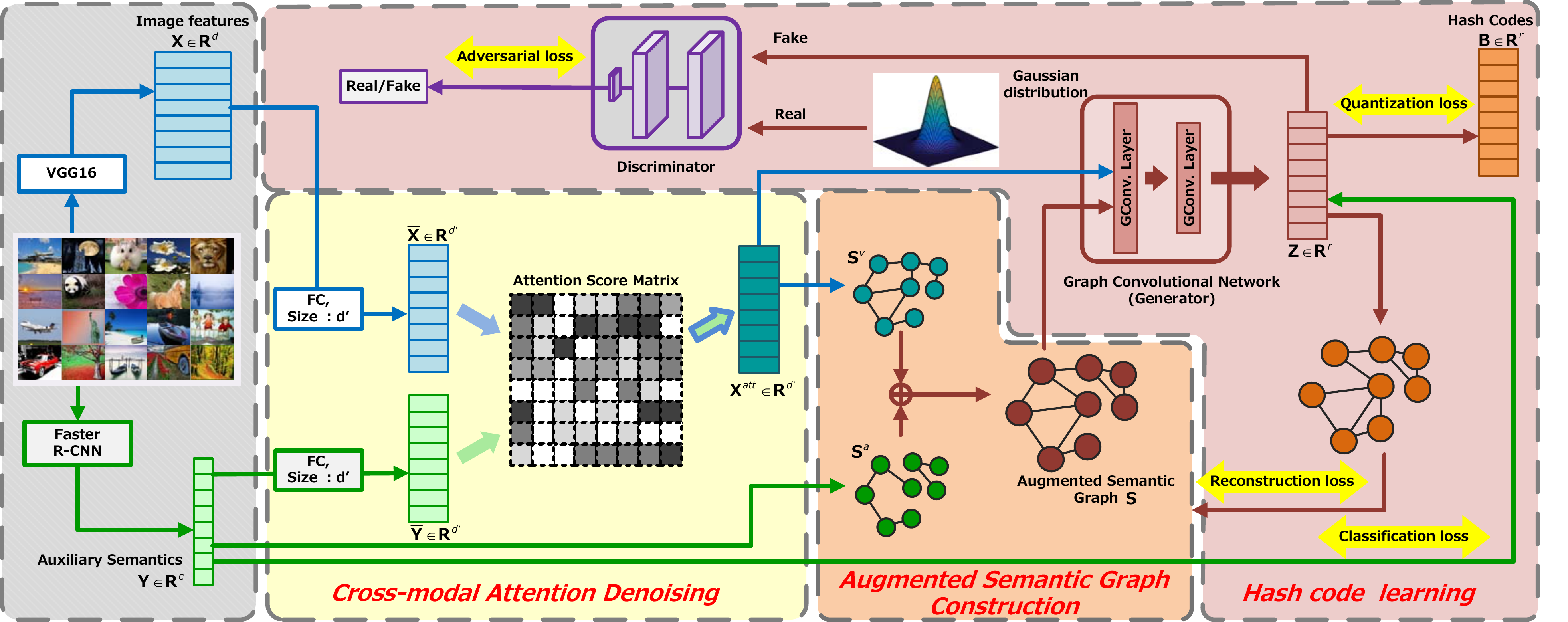}}
\caption{The basic learning framework of the proposed LAGNH method. We first extract the deep image features with the pre-trained VGG16 model \cite{VGG16} and detect the auxiliary semantics from the inner structure of the image with Faster R-CNN \cite{fasterrcnn2017}. Then, a cross-modal attention module is designed to identify the important parts of the images that are highlighted by the auxiliary semantic information. The attentive image features and auxiliary semantics are integrated together to construct an augmented semantic graph. Afterwards, an adversarial regularized graph convolutional network is proposed to learn the hash codes by reconstructing the semantic relations of images.}
\vspace{-4mm}
\label{figfram}
\end{figure*}

In this paper, we propose a \emph{Lightweight Augmented Graph Network Hashing} (LAGNH) method to address the problems with an effective two-pronged strategy. Our model takes full advantage of the free auxiliary semantics extracted from the inner structure of the image to improve both the semantic representation capability of binary hash codes and model training efficiency. For one thing, we extract the auxiliary semantics and employ them as the supervision to guide the training process of hash learning, which overcomes the shortage of semantic supervision in unsupervised deep hashing. In particular, we first design a cross-modal attention module with the auxiliary semantic information to adaptively mitigate the noise in the deep image features. Then, the attentive image features and the auxiliary semantics are used together to construct the augmented semantic graph, which captures the intrinsic semantic relations of images. Afterwards, the hash codes are learned by an adversarial regularized graph convolutional network based on the attentive image features and the augmented semantic graph. For another, with the assistance of the auxiliary semantics, a lightweight attention module and a graph convolutional network are constructed, which avoids the over-abundant parameters in deep networks. Moreover, we reconstruct a similarity graph to further transfer the captured semantic relations of images into the hash codes. Via this two-pronged strategy, the retrieval accuracy and training efficiency of our unsupervised deep hashing can be improved simultaneously. Figure \ref{figfram} illustrates the basic learning framework of our approach. We highlight the contributions of this paper as follows:
\begin{itemize}
\item We propose a two-pronged strategy to address the two key problems in existing unsupervised deep hashing methods. For one thing, we extract the free auxiliary semantics and employ them as the supervision to guide the training process of hash learning. For another, with the assistance of auxiliary semantics, we design a lightweight network to avoid the over-abundant parameters in deep networks and thus accelerate the training process. To the best of our knowledge, there is still no similar work. \vspace{1mm}
\item Technically, we design a cross-modal attention module based on the extracted auxiliary semantic information to adaptively mitigate the adverse effects in the deep image features. Besides, an adversarial regularized graph convolution network based on the denoised features and the augmented semantic graph is proposed to learn the semantically enhanced hash codes by reconstructing the captured semantic relations. \vspace{1mm}
\item Experimental results on two public image retrieval datasets demonstrate that our LAGNH method is superior to the baseline methods in terms of both retrieval accuracy and training efficiency. Notably, on MS-COCO dataset, our method achieves more than 10\% improvement on retrieval precision and 2.7x speedup on training time compared with the second best result.
\end{itemize}
\vspace{-2mm}
\section{Related Work}
Unsupervised hashing methods learn hash functions without the reliance on the labeled data. They can be further divided into two categories: shallow methods and deep methods.

\textbf{Shallow learning methods}. Locality Sensitive Hashing (LSH) \cite{LSH1999} and Shift-Invariant Kernels Hashing (SKLSH) \cite{SKLSH2009} directly map data to a low-dimensional subspace using random projections. Spectral Hashing (SH) \cite{SH2008} performs spectral analysis on high-dimensional data, and transforms the binary hash coding problem into a dimension reduction problem of the graph Laplacian by relaxing the constraints. Anchor Graph Hashing (AGH) \cite{AGH2011} constructs anchor graphs for solving the eigenfunctions of the graph Laplacians, which improves the efficiency of graph hashing. Principal Component Analysis Hashing (PCAH) \cite{PCAH2012} employs principal component analysis to learn a hash projection matrix. Iterative Quantization (ITQ) \cite{ITQ2013} first projects data into a low-dimensional space by principal component analysis, and then finds an orthogonal transformation matrix to minimize the quantization loss. Scalable Graph Hashing (SGH) \cite{SGH2015} approximates the similarity graph by feature transformation, and proposes a strategy to learn hash functions in a bit-wise manner. Latent Semantic Minimal Hashing (LSMH) \cite{LSMH2017} refines the original features by matrix factorization, and minimizes the encoding loss with an orthogonal transformation to learn discriminative hash codes.

\textbf{Deep learning methods}. Recently, inspired by the success of deep learning, unsupervised deep hashing methods are proposed to jointly perform image representation learning and hash function learning using deep neural networks. Several unsupervised deep hashing methods have been proposed. Unsupervised Hashing with Binary Deep Neural Network (UH-BDNN) \cite{UH-BDNN2016} regards the output of the penultimate layer of the designed neural network as hash codes, and enforces the desirable properties of hash codes during the learning process: similarity preserving, independence and balance. DeepBit \cite{DeepBit2016} introduces a deep convolutional neural network and enforces three important criteria on the top layer of the network to learn hash functions and discriminative binary codes. Semantic Structure-based Deep Hashing (SSDH) \cite{SSDH2018} constructs a pair-wise semantic similarity structure by analyzing the data distribution of deep features. Then, the learned similarity structure is integrated into the deep learning architecture to obtain the similarity-preserving hash codes. Similarity-Adaptive Deep Hashing (SADH) \cite{SADH2018} alternately trains three modules in one deep learning framework. The preceding module guides the subsequent module, which makes the hash codes to be more compatible with the deep hash functions. Twin-Bottleneck Hashing (TBH) \cite{TBH2020} designs an auto-encoding twin-bottleneck: a binary bottleneck and a continuous bottleneck. The former explores the underlying data structure by adaptively constructing code-driven similarity graphs, while the latter adopts data relevance information from the binary codes for high-quality decoding and reconstruction.

The aim of deep hashing is to exploit the powerful representation capability of deep networks. However, since a large number of parameters are involved in the deep neural networks and no explicit semantic supervision can be exploited to guide the hash learning process, the current unsupervised deep hashing models are very difficult to optimize and their learning processes suffer from low training efficiency. The retrieval accuracy and training efficiency of existing unsupervised deep hashing are still limited. Different from existing methods, in this paper, we propose a two-pronged strategy to address the problems. For one thing, we overcome the shortage of semantic supervision in unsupervised deep hashing by extracting the free auxiliary semantics and employing them as the supervision to guide the training process of deep hash learning. For another, we design a lightweight network structure with the assistance of auxiliary semantics, which greatly reduces the number of network parameters that need to be optimized and thus greatly accelerates the training process. To the best of our knowledge, there is still no similar work.
\section{The Proposed Method}\label{method}
Assuming that a training set contains $n$ images,
our goal is to learn the hash functions which project the input images into compact binary codes
$\mathbf{B} = \left[ {\mathbf{b}_1,\cdot \cdot \cdot ,\mathbf{b}_n} \right] \in {\left\{ { - 1,1} \right\}^{r \times n}}$,
where $r$ is the length of the hash code.
We denote the deep image features as $\mathbf{X}= \left[ {\mathbf{x}_1,\cdot \cdot \cdot,\mathbf{x}_n} \right] \in {\mathbb{R}^{d \times n}}$,
where $d$ is the dimensionality of each image vector.
The auxiliary semantic information extracted from the inner structure of the image is denoted as $\mathbf{Y}= \left[{\mathbf{y}_1,\cdot \cdot \cdot,\mathbf{y}_n} \right] \in {\mathbb{R}^{c \times n}}$,
where $c$ is the number of semantic categories.
$\mathbf{y}_i$ is a binary-valued vector, in which $y_{ij}=1$ indicates that the $j$-th semantic label is associated with the $i$-th image,
and $y_{ij}=0$ otherwise.
\vspace{-2mm}
\subsection{Framework Overview}
Figure. \ref{figfram} shows the basic learning framework of LAGNH. It mainly consists of four components:
\begin{itemize}
  \item \emph{Preparation}. The deep image features and auxiliary semantic information are extracted by deep learning models.
  \item \emph{Cross-modal Attention Denoising}. Identifying the important parts of the image features that are highlighted by the auxiliary semantic information and mitigating the adverse effects in the features.
  \item \emph{Augmented Semantic Graph Construction}. Integrating the attentive image features and auxiliary semantics to construct the augmented semantic graph for preserving semantic relations.
  \item \emph{Hash Code Learning}. Learning semantically enhanced hash codes by an adversarial regularized graph convolution network and a similarity graph reconstruction strategy.
\end{itemize}
Below, we elaborate our method from the aspects of cross-modal attention denoising, augmented semantic graph construction, and hash learning based on the adversarial regularized graph convolutional network.
\vspace{-2mm}
\subsection{Cross-modal Attention Denoising}
Deep hashing takes advantage of the representation capability of deep neural networks to obtain features that can capture the global visual information in images. However, the deep image features contain noise, which needs to be removed to alleviate the adverse effects on hashing performance.
Since each image is more semantically relevant to specific auxiliary semantic information, we develop a cross-modal attention module \cite{add7,add9} to identify the important parts of the image that are highlighted by the auxiliary semantic information. Specifically, for each image feature ${{\mathbf{x}_i}}$ in the training set, we calculate the cosine similarities with all the auxiliary semantics $\left\{ {{\mathbf{y}_j}} \right\}_{j = 1}^n$ to get the attention score matrix $\mathbf{A}^{att} \in {\mathbb{R}^{n \times n}}$. Each item in $\mathbf{A}^{att}$ is calculated as below:
\begin{equation}
{\alpha _{ij}} =  \left[ {\cos ({\bar {\mathbf{x}}_i},{\bar {\mathbf{y}}_j})} \right]_+,
\end{equation}
where ${\bar {\mathbf{x}}_i} \in {\mathbb{R}^{d^\prime}} $ and ${\bar {\mathbf{y}}_j} \in {\mathbb{R}^{d^\prime}}$
are the linearly transformed representation of ${\mathbf{x}}_i$ and $ {\mathbf{y}}_j$, respectively.
$\cos(\cdot , \cdot)$ returns the cosine value of the input vectors and
the operator ${\left[ \alpha \right]_ + } = \max \left( {\alpha,0} \right)$.
Then the attentive feature of ${\mathbf{x}}_i$ is calculated
by adding the weighted sum of the auxiliary semantic information %using the attention scores
to the linearly transformed representation ${\bar {\mathbf{x}}_i}$, namely,
\begin{equation}
\mathbf{x}_i^{att} = \frac{{\sum\nolimits_{j = 1}^n {{\alpha _{ij}}{{\bar {\mathbf{y}}}_j}} }}{{\sum\nolimits_{j = 1}^n {{\alpha _{ij}}} }} + {\bar {\mathbf{x}}_i} .
\end{equation}
\vspace{-2mm}
\subsection{Augmented Semantic Graph Construction}
To enhance the semantic representation capability of the deep network, we jointly consider the deep image features and the auxiliary semantic information. The visual similarity based on the attentive deep image features can capture the relations between images at a coarse-grained level. The similarity based on auxiliary semantics extracts the relations from the inner structures of the images, and it can discover the fine-grained semantic association. Thus, the visual similarity and the auxiliary semantic similarity are complementary to each other. By integrating the two forms of similarities together, we construct the augmented semantic graph as the supervision to guide our hash model training.

We use the Gaussian kernel function to calculate the visual similarity graph. For any two images, such as image $i$ and image $j$, the visual similarity can be described with the following formula:
\begin{equation}\label{eq01}
S_{ij}^{v} = \exp \left( {\frac{{ - {{\left\| {\mathbf{x}_i^{att} - \mathbf{x}_j^{att}} \right\|}^2}}}{2 \sigma ^2 }} \right),
\end{equation}
where $\sigma$ is the kernel bandwidth.

Following the previous studies \cite{DCMH2017,ADSH2018}, the inner product of $\mathbf{y}_i$ and $\mathbf{y}_j$ is regarded as the auxiliary similarity between image $i$ and $j$, that is
\begin{equation}\label{eq02}
S_{ij}^{a}=\left({\mathbf{y}_i}\right)^\texttt{T}{\mathbf{y}_j}.
\end{equation}
Ultimately, we get the augmented semantic graph $\mathbf{S}$ which employs the information of both image features and auxiliary semantics as follows:
\begin{equation}\label{eq04}
\mathbf{S} = \mu{\mathbf{S}}^{v} + {\mathbf{S}}^{a},
\end{equation}
where $\mu$ is the fusion factor to balance the importance of each similarity structure. At this point, the intrinsic semantic relations of images are captured and modeled.
\vspace{-2mm}
\subsection{Hash Code Learning}
The goal of hashing is to project the images into binary hash codes while preserving their semantic similarities. Recent studies have demonstrated that Graph Convolutional Networks (GCNs) are capable of generating similarity-preserving node representations based on the node features and topological structure of the graph \cite{GCN2017,GAE2017,ARGA2018,add1,zhu6}. In this paper, with the semantic assistance of auxiliary semantics, we introduce a lightweight two-layer GCN to fuse the feature of the node in the augmented semantic graph $\mathbf{S}$ with its neighbor nodes. The propagation rules of each graph convolutional layer can be represented as
\begin{align}
&{\mathbf{Z}^{\left( 1 \right)}} = \sigma_{1} \left( {\mathbf{W}^{\left( 1 \right)}} \mathbf{X}^{att} { \mathbf{\tilde S}} \right), \\
&{\mathbf{Z}^{\left( 2 \right)}} = \sigma_{2} \left( {\mathbf{W}^{\left( 2 \right)}}{\mathbf{Z}^{\left( 1 \right)}} { \mathbf{\tilde S}} \right).
\end{align}
$\mathbf{\tilde S} = {\mathbf{D}^{ - 1/2}}\mathbf{S}{\mathbf{D}^{ - 1/2}} \in {\mathbb{R}^{n \times n}}$ is the normalized augmented semantic graph,
where ${\mathbf{D}_{ii}} = \sum\nolimits_j {\mathbf{S}_{ij}}$.
${\mathbf{W}^{\left( 1 \right)}}$ and ${\mathbf{W}^{\left( 2 \right)}}$ are the matrices of filter parameters, which need to be learned during training.
$\sigma_{l} \left(  \cdot  \right)$ denotes the activation function of the $l$-th layer.
${\mathbf{Z}}^{\left( 2 \right)}$ represents the outputs of our GCN. We replace it with ${\mathbf{Z}}$ for simplicity.

In essence, $\mathbf{Z}$ aggregates the semantic relationships from the graph structure and node features. Based on it, we try to minimize the gap between the hash codes and the output node representations of the GCN, that is
\begin{equation}\label{lossquan}
\min \mathcal L_{quan} = \left\| {\mathbf{B} - \mathbf{Z}} \right\|_\texttt{F}^2.
\end{equation}

After that, instead of reconstructing the original images, we heuristically employ the cosine similarity of the outputs of the GCN to reconstruct the auxiliary similarity graph. The reason why we reconstruct the auxiliary similarity graph instead of the augmented semantic graph will be explained in the experiments section. We formulate this part as
\begin{equation} \label{lossrecons}
\min \mathcal L_{recons} = {\left\| {k{\mathbf{S}^{a}} - [\cos ({{\mathbf{Z}}^{\texttt{T}}},{\mathbf{Z}})}]_+ \right\|_\texttt{F}^2},
\end{equation}
where $k$ is the hyper-parameter to make the reconstruction more flexible. The new representations can be viewed as the continuous representations of the hash codes, so the similarities between images are transferred into the hash codes. Eq. (\ref{lossrecons}) works as follows: the cosine similarities between the node representations can reflect the angular relations of hash codes in Hamming space, and these angular relations are equivalent to the Hamming distance of hash codes.

In addition, we enforce the data distribution of the new representations to match the real data distribution. This part is implemented by Generative Adversarial Nets (GANs) \cite{GAN2014}. A discriminator is designed to distinguish whether a representation is from the real data distribution (positive) or from the outputs of the GCN (negative), and our GCN is regarded as the generator. The objective function of the adversarial training strategy is formulated as
\begin{equation} \label{lossgan}
\begin{aligned}
\mathop {\min }\limits_{\mathcal{G}} \mathop {\max }\limits_{\mathcal{D}} \mathcal L_{GAN} =  &\mathbb{E}_{\mathbf{z}^\prime \sim p_{\mathbf{z}^\prime}}[\log \mathcal{D}(\mathbf{Z}^\prime)] \\
+ &\mathbb{E}_{\mathbf{z}\sim p_{\mathbf{z}}}[\log (1 - \mathcal{D}(\mathcal{G}(\mathbf{Z})))],
\end{aligned}
\end{equation}
where $\mathcal{G}\left(  \cdot  \right)$ and $\mathcal{D}\left(  \cdot  \right)$ indicate the generator and discriminator respectively,
$\mathbf{z}^\prime$ is sampled from the real data distribution.
In our paper, we use simple Gaussian distribution as $p{(\mathbf{z}^\prime)}$. This training scheme will regularize the outputs of the GCN to be more robust graph representations.

Furthermore, as we have obtained the auxiliary semantic information, it is natural to consider using it to directly guide our hash function learning, which can be represented as
\begin{equation} \label{losscl}
\mathop {\min }\limits_{\mathbf{Z}} \mathcal L_{cl} = \sum\limits_{i = 1}^n {{\mathbf{y}_i}} \log (\delta (\mathbf{z}_i)) + (1 - {\mathbf{y}_i})\log (1 - \delta (\mathbf{z}_i)),
\end{equation}
where $\delta (\mathbf{z}_i) = \frac{1}{1 + {e^{ - {\mathbf{z}_i}}}}$.

By combining Eq. (\ref{lossquan}) - Eq. (\ref{losscl}), the overall objective function of our hash learning framework is given as
\begin{equation} \label{losstotal}
\mathcal L = \mathcal L_{GAN} + \lambda_1 \mathcal L_{recons} +\lambda_2 \mathcal L_{quan} + \lambda_3 \mathcal L_{cl},
\end{equation}
where $\lambda_1,\lambda_2,\lambda_3$ are tradeoff parameters.
\begin{table}
%\footnotesize
\caption{Network configuration of the proposed hash code learning module, where ${d^\prime}$ denotes the dimensionality of ${\bar {\mathbf{x}}}$ or ${\bar {\mathbf{y}}}$ , $r$ denotes the length of hash codes. Note that we set the output dimension of GC Layer 1 to 1024 on MS COCO and 2048 on NUS-WIDE, respectively.}
\setlength{\tabcolsep}{5mm}
\begin{tabular}{l|c|c}
\toprule
Layer                   & Configuration       & Activation    \\ \hline \hline
\multicolumn{3}{l}{\textbf{Graph Convolutional Network (Generator)}} \\ \hline
GC Layer 1     &  ${d^\prime} \times$ 1024/2048     & ReLU          \\ \hline
GC Layer 2     & 1024/2048 $\times r  $     & -             \\ \hline
\multicolumn{3}{l}{\textbf{Discriminator}}                           \\ \hline
FC1                     & $r\times$64                & ReLU          \\ \hline
FC2                     & 64$\times$32               & ReLU          \\ \hline
FC3                     & 32$\times$1                & -             \\ \bottomrule
\end{tabular} \label{VGAN}
\end{table}

We summarize the network configuration of our hash code learning module in Table \ref{VGAN}, which contains a two-layer GCN and a three-layer discriminator. As can be seen from the table, our hash learning model has a simple network structure and only few parameters, so it is easy to be optimized and thus can accelerate the training process. We will demonstrate the advantage of our proposed method on training efficiency in Subsection \ref{time}.

After several iterative optimizations, our model is fully trained and the optimal network parameters are obtained.
For any query image $q$, we can apply its auxiliary semantic information and the trained model parameters to generate the binary code.
Specifically, we first get the low-dimensional representation $\mathbf{z}_q$ of our GCN by forward propagation.
Then, the hash code $\mathbf{b}_q$ is obtained by a simple quantization, that is
\begin{equation}
\mathbf{b}_q =  \text{sgn} (\mathbf{z}_q).
\end{equation}
sgn$(\cdot)$ denotes the element-wise sign function which returns $1$ if the element is positive and returns $-1$ otherwise.
\begin{table}
%\footnotesize
\caption{Statistics of the experimental data}
\vspace{-3mm}
\centering
\begin{tabular}{lccc}
  \toprule
  Datasets & MS COCO &  NUS-WIDE \\
  \hline \hline
  \#Categories & 80      & 21      \\
  \#Training   & 5,000   & 5,000   \\
  \#Query      & 5,000   & 3,156   \\
  \#Retrieval  & 113,287 & 137,577   \\
  \bottomrule
\end{tabular}
\label{table2}
\end{table}
\section{Experiments}
\subsection{Datasets and Evaluation Metrics}
We conduct extensive experiments to evaluate the performance of the proposed method on two public image retrieval datasets. These two datasets have been widely evaluated in recent unsupervised deep hashing methods \cite{MLS3RDUH2020,TBH2020,BGAN2018,SADH2018}.

\textbf{MS COCO} \cite{COCO} is a large-scale dataset provided by the Microsoft team for object detection, segmentation, and captioning. It contains 118,287 training images and 5,000 validation images, each of which is associated with at least one of 80 object categories. In our experiments, we randomly select 5,000 images from the provided training data as our training set, and the remaining images as our retrieval set. The provided validation data is directly used as our query set.

\textbf{NUS-WIDE} \cite{NUSWIDE} is the largest publicly available dataset used for evaluating image retrieval performance. Each image in the dataset is associated with user-provided tags.  We first select images that fall into the 21 most common tags. Then, we remove the images that do not belong to the 21 most common categories and obtain 145,733 experimental images. We randomly select 200 images from each category, and finally obtain the query set with 3,156 images after removing the duplicate ones. For the remaining images, 5,000 images are selected as the training set, and 137,577 images are selected as the retrieval set. Table \ref{table2} summarizes the basic statistics of the datasets in our experiments.

Following the settings as existing hashing methods \cite{UDDH2017,lxMM2019,MGRN2020,add5,add6}, we use Mean Average Precision (MAP) and topK-Precision curve as the evaluation metrics in our experiments. Since both of the above testing datasets are multi-labeled, two images are deemed as semantically similar if they share at least one label (category). For the evaluation metrics, higher values indicate better outcomes. Note that the label information is only used for evaluating the retrieval performance, and it is not used during the training process.
\vspace{-2mm}
\subsection{Implementation Details}
Our LAGNH method is implemented on Pytorch\footnote{https://pytorch.org/} platform (an open source machine learning framework).
All the experiments are performed on a workstation with one NVIDIA GeForce GTX 1080Ti GPU and 64 GB memory.
Our network is trained by the Adam optimizer \cite{Adam}, and the learning rate is set to 0.0001 in optimization.
The batchsize is set to 1024 for MS COCO and 2048 for NUS-WIDE at the training stage.

We use the VGG16 model \cite{VGG16,add8} pre-trained on ImageNet dataset \cite{imagenet} as our feature extractor.
The 4,096-dimensional deep image features are acquired from the last fully connected layer (FC7) of VGG16.
The computer vision model Faster R-CNN \cite{fasterrcnn2017} is used to parse the inner structures of the images,
where the identified objects with high prediction scores are taken as the auxiliary semantic information of MS COCO.
As Faster R-CNN is not applicable to NUS-WIDE, we employ the available user-provided tags directly as an alternative.
%Our discriminator $\mathcal{D}$ is implemented by a 3-layer feed-forward neural network whose dimension of output layer is 1.

The hyper-parameters are determined by cross-validation and are finally set as
% beta = lambda_1, lamda = \lambda_2, \alpha =\lambda_3
$\{\lambda_1=100$, $\lambda_2=1$, $\lambda_3=1$, $k=1$, $\mu=1\}$
and $\{\lambda_1=100$, $\lambda_2=1$, $\lambda_3=0.1$, $k=1$, $\mu=1\}$ on MS COCO and NUS-WIDE, respectively.
The whole hash learning framework is optimized for 300 epochs in total.
\begin{table*}
%\footnotesize
\centering
\caption{{MAP@1000 results on MS COCO and NUS-WIDE.}
The best result in each column is marked with bold.
The second best result in each column is underlined.
}
\vspace{-3mm}
%\scriptsize
\begin{tabular}{llcccccccc}
\toprule
\multirow{2}{*}{Methods}&\multirow{2}{*}{Reference}& \multicolumn{4}{c}{MS COCO}           & \multicolumn{4}{c}{NUS-WIDE} \\ \cline{3-10}
                         && 16 bits & 32 bits & 64 bits & 128 bits & 16 bits & 32 bits & 64 bits & 128 bits \\ \hline\hline

VGG+LSH \cite{LSH1999}     &VLDB99  &0.4548 	&0.5474 &0.6310 &0.7149 &0.4777	&0.5314 &0.6066 &0.6761 	\\
VGG+SKLSH \cite{SKLSH2009} &NeurIPS09  &0.3847 	&0.4622 &0.5016 &0.6029 &0.4250	&0.4632 &0.5146 &0.5633 	\\
VGG+SH \cite{SH2008} 	  &NeurIPS08  &0.6654 	&0.7055 &0.7321 &0.7482 &0.6249	&0.6393 &0.6667 &0.6807 	\\
%VGG+AGH-1\cite{AGH2011}  &ICML11  &0.7071 	&0.7601 &0.7806 &0.7697 &0.6780	&0.7053 &0.7160 &0.7184 	\\
%VGG+AGH-2\cite{AGH2011}  &ICML11  &0.6489 	&0.7407 &0.7794 &0.7966 &0.6719	&0.6954 &0.7142 &0.7236 	\\
VGG+PCAH \cite{PCAH2012}   &TPAMI12 &0.6619 	&0.7054 &0.7251 &0.7300 &0.6338	&0.6490 &0.6615 &0.6774 	\\
VGG+ITQ \cite{ITQ2013} 	  &TPAMI13 &0.6920 	&0.7607 &0.7895 &0.8067 &\underline{0.6705}	&\underline{0.6976} &\underline{0.7218} &0.7368 	\\
VGG+SGH \cite{SGH2015}  	  &IJCAI15 &0.6769 	&0.7416 &0.7716 &0.7929 &0.6576	&0.6833 &0.7083 &0.7250 	\\
%VGG+LSMH\cite{LSMH2017}  &TIP17   &0.6927 	&0.7590 &0.7907 &0.8061 &0.6996	&0.6992 &0.7216 &0.7361 	\\
\hline % deep
VGG+UH-BDNN \cite{UH-BDNN2016}&ECCV16  &0.6866 &0.7472 &0.7430 &0.7789 &0.6678 &0.6845 &0.6770 &0.7037  	\\
Deepbit \cite{DeepBit2016}&CVPR16  &0.4649 &0.5360 &0.6346 &0.7120 &0.4552 &0.5464 &0.6032 &0.6733  	\\
SSDH \cite{SSDH2018}	     &IJCAI18 &0.4353 &0.4667 &0.4619 &0.5319 &0.5303 &0.5339 &0.5352 &0.5311 	\\
SADH \cite{SADH2018}	     &TPAMI18 &0.6728 &\underline{0.7605} &\underline{0.7981} &\underline{0.8161} &0.6506 &0.6892 &0.7215 &\underline{0.7436} 	\\
TBH \cite{TBH2020}	     &CVPR20  &\underline{0.6942} &0.7134 &0.7299 &0.6912 &0.6560 &0.6780 &0.6791 &0.6981 	\\
\hline									
Ours (LAGNH)	&Proposed &\textbf{0.8476} &\textbf{0.8655} &\textbf{0.8718} &\textbf{0.8751} &\textbf{0.7061} 	&\textbf{0.7242} 	&\textbf{0.7530} &\textbf{0.7644} 	\\
\bottomrule
\end{tabular}\label{mapresults}
%\vspace{-5mm}
\end{table*}
\begin{figure*}
\centering
\subfigure{\includegraphics[scale=0.29]{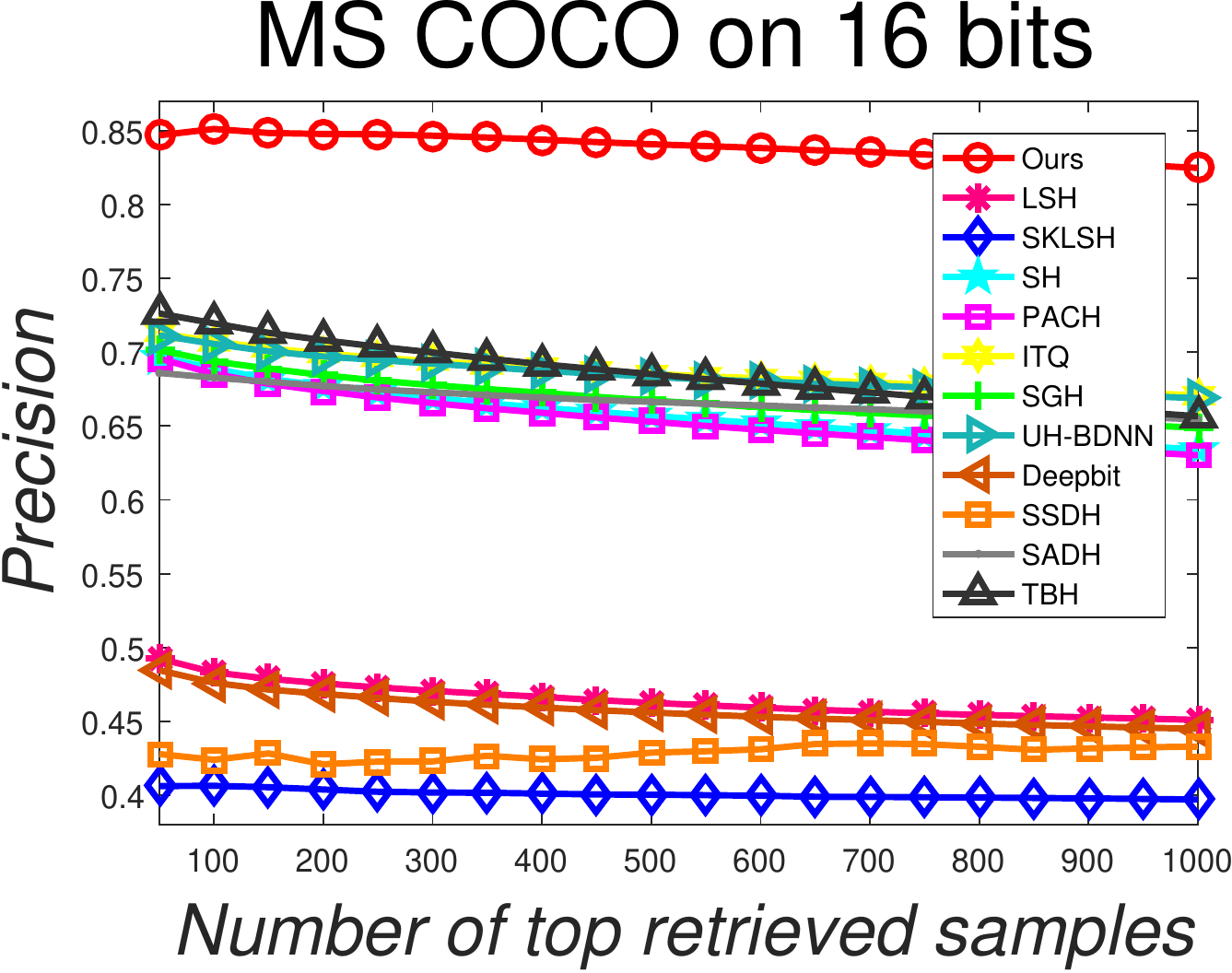}}
\subfigure{\includegraphics[scale=0.29]{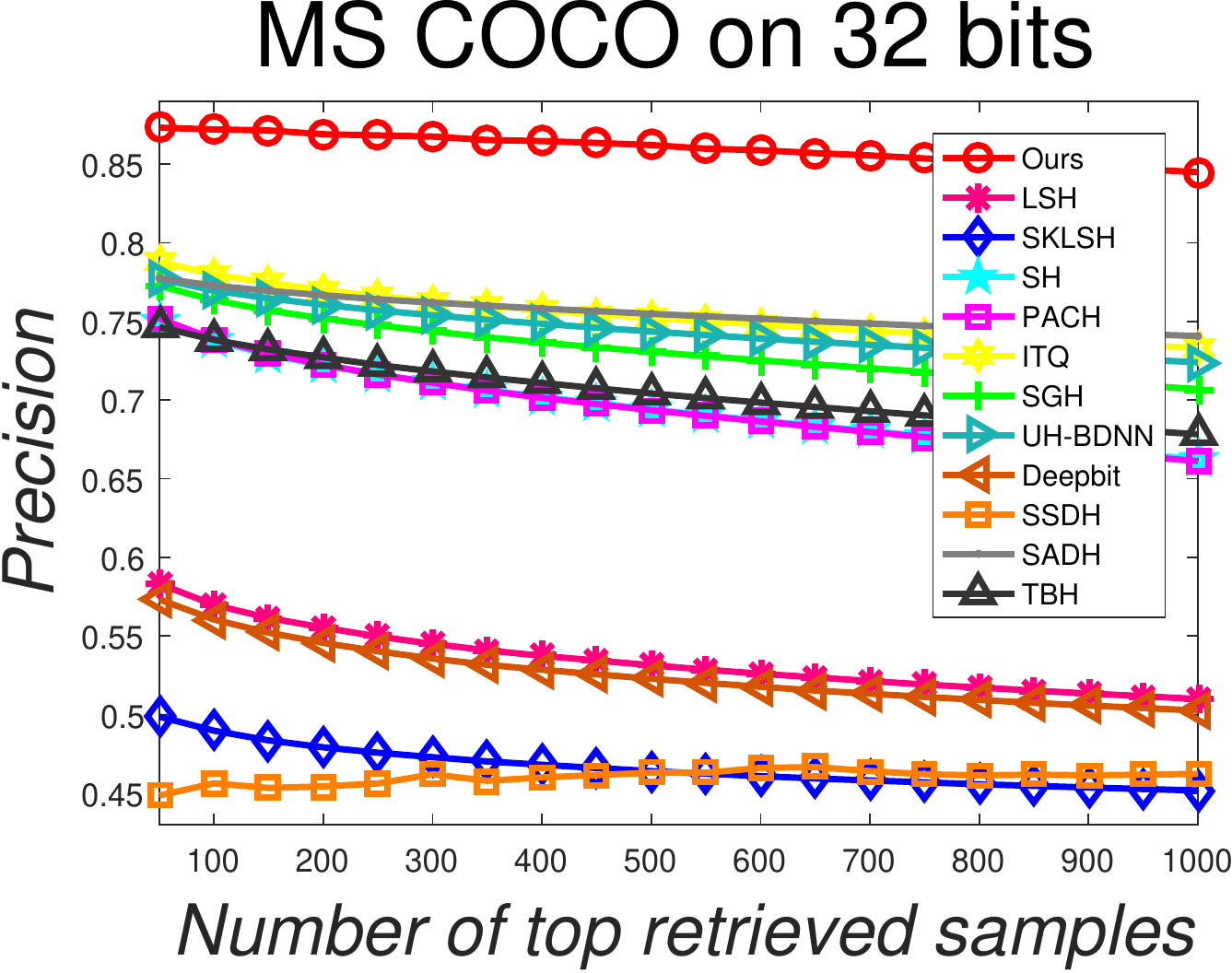}}
\subfigure{\includegraphics[scale=0.29]{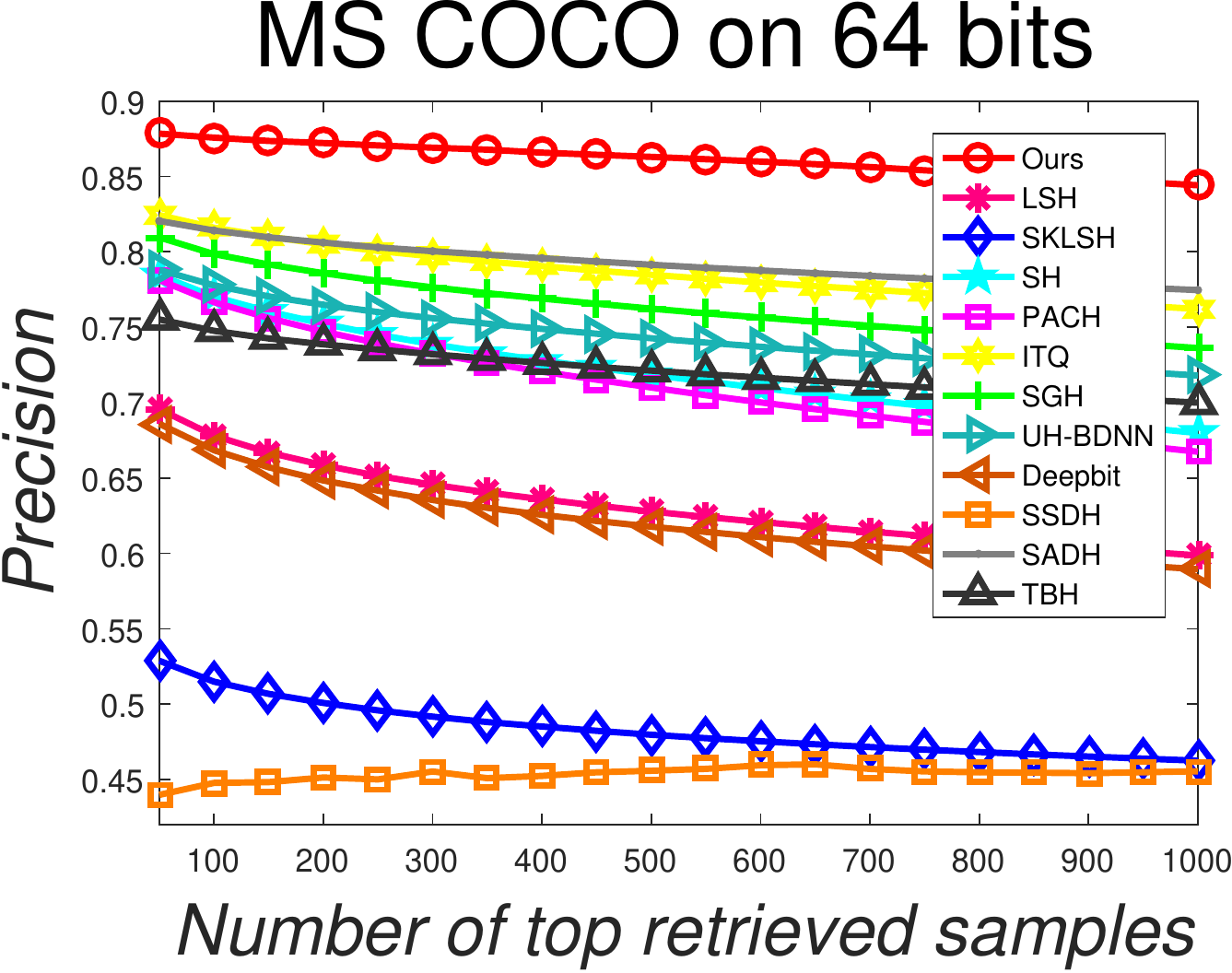}}
\subfigure{\includegraphics[scale=0.29]{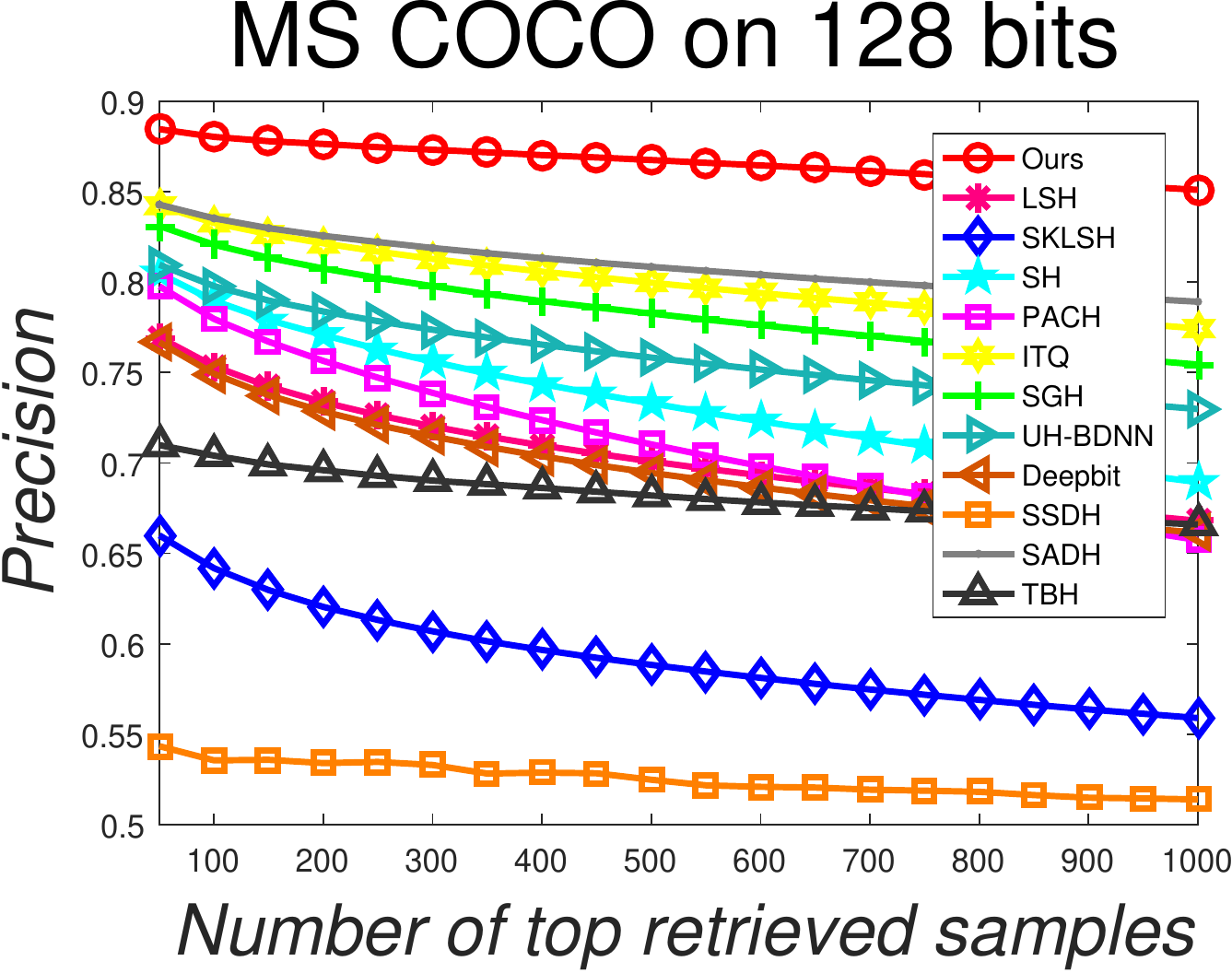}}\\
\subfigure{\includegraphics[scale=0.29]{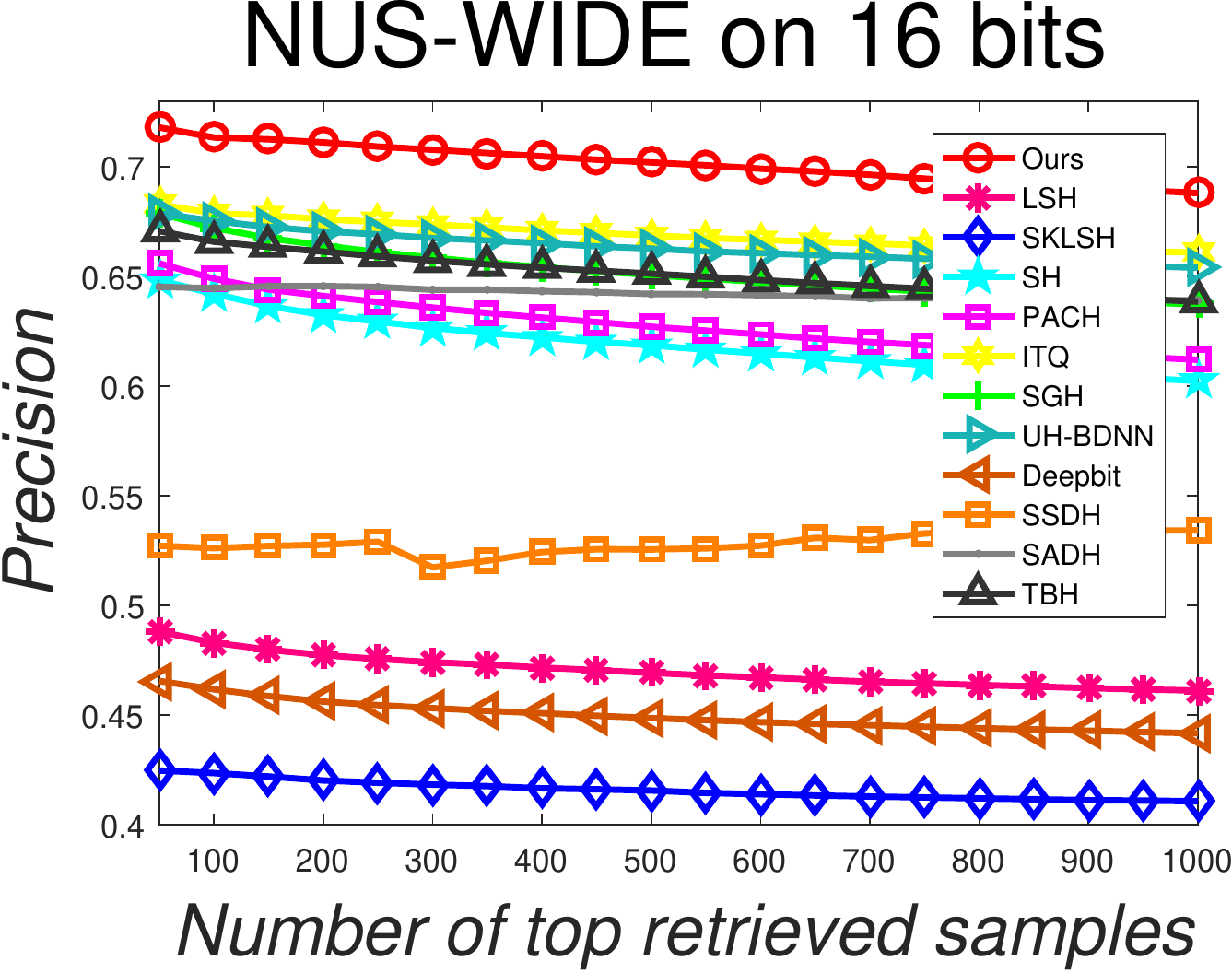}}
\subfigure{\includegraphics[scale=0.29]{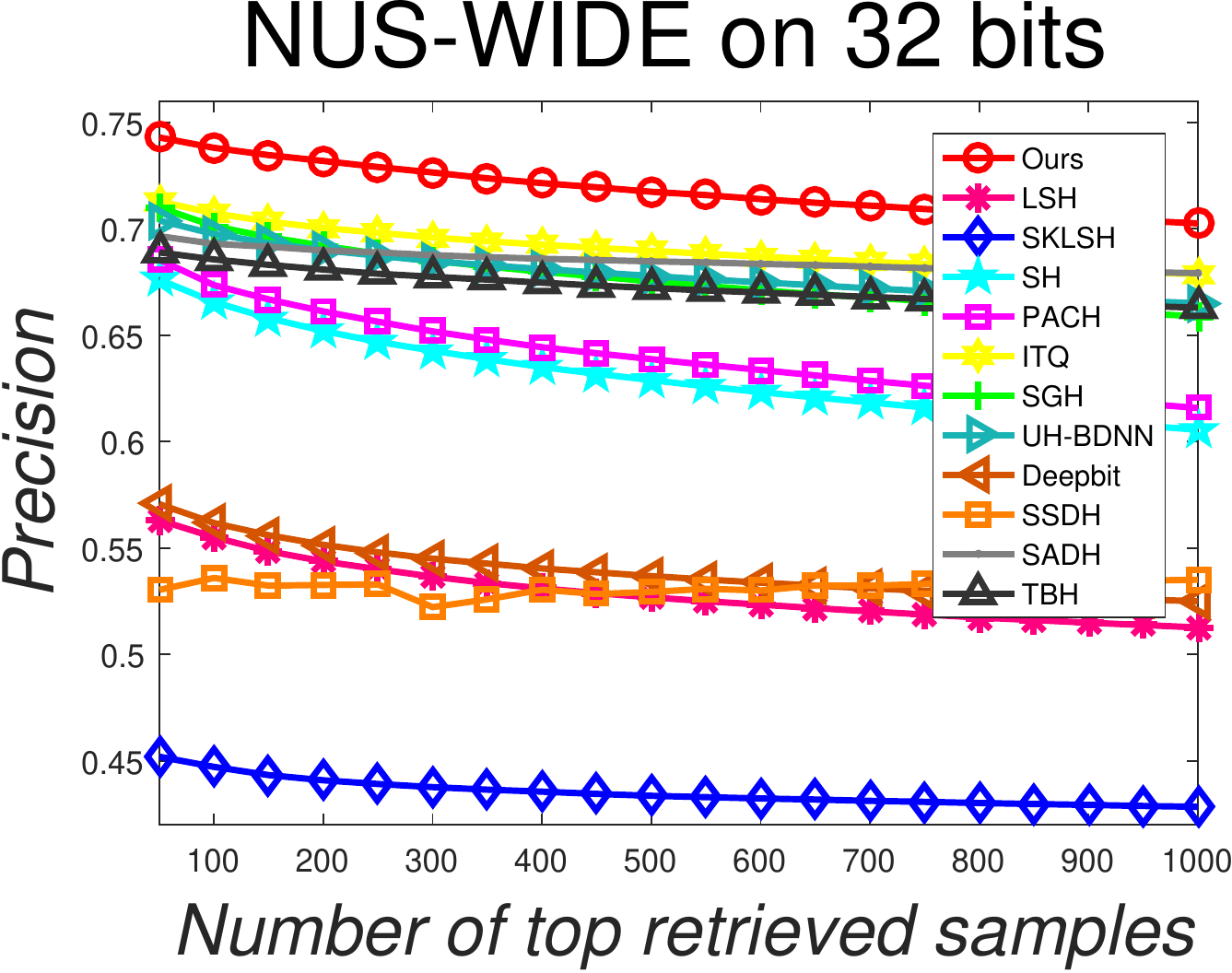}}
\subfigure{\includegraphics[scale=0.29]{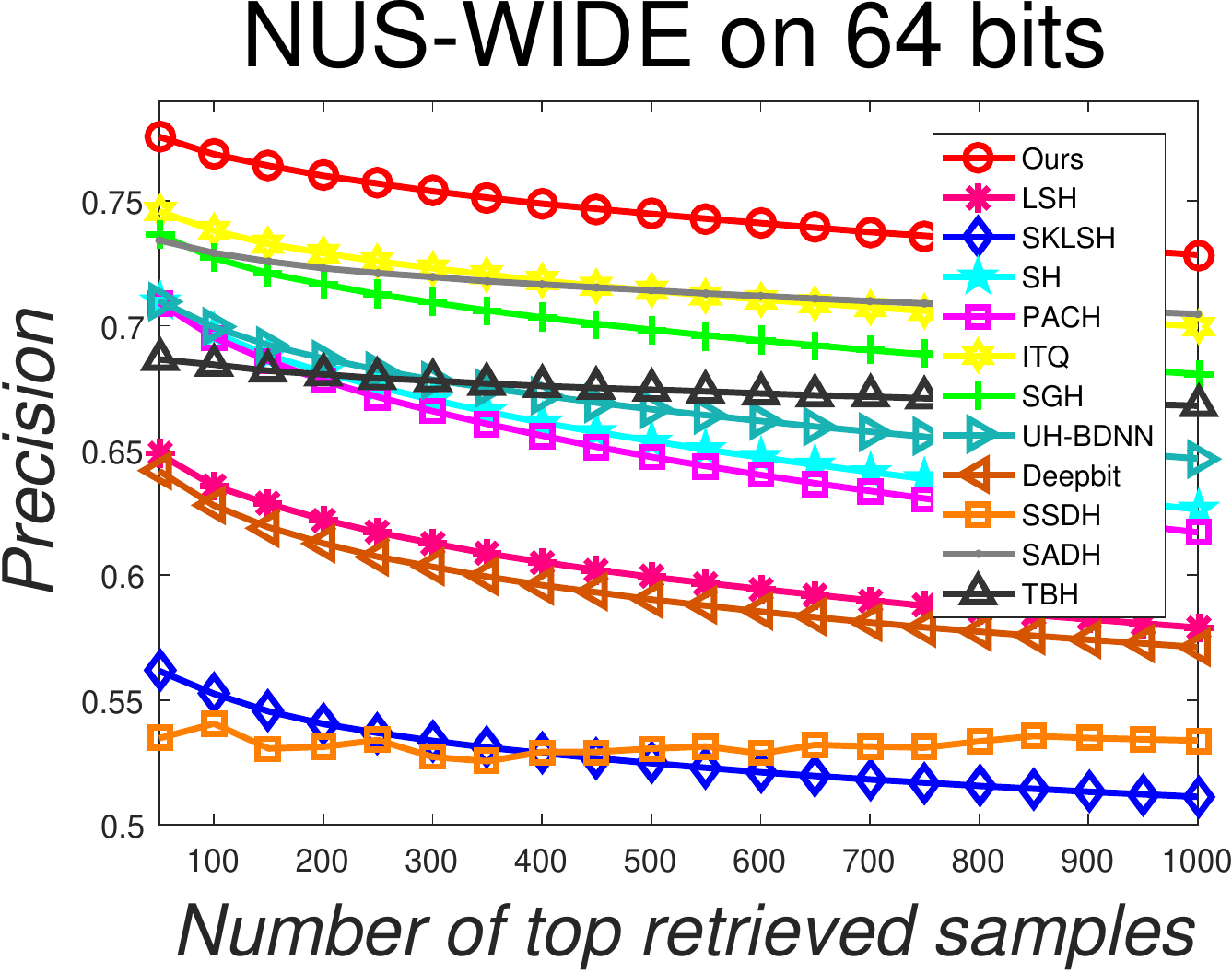}}
\subfigure{\includegraphics[scale=0.29]{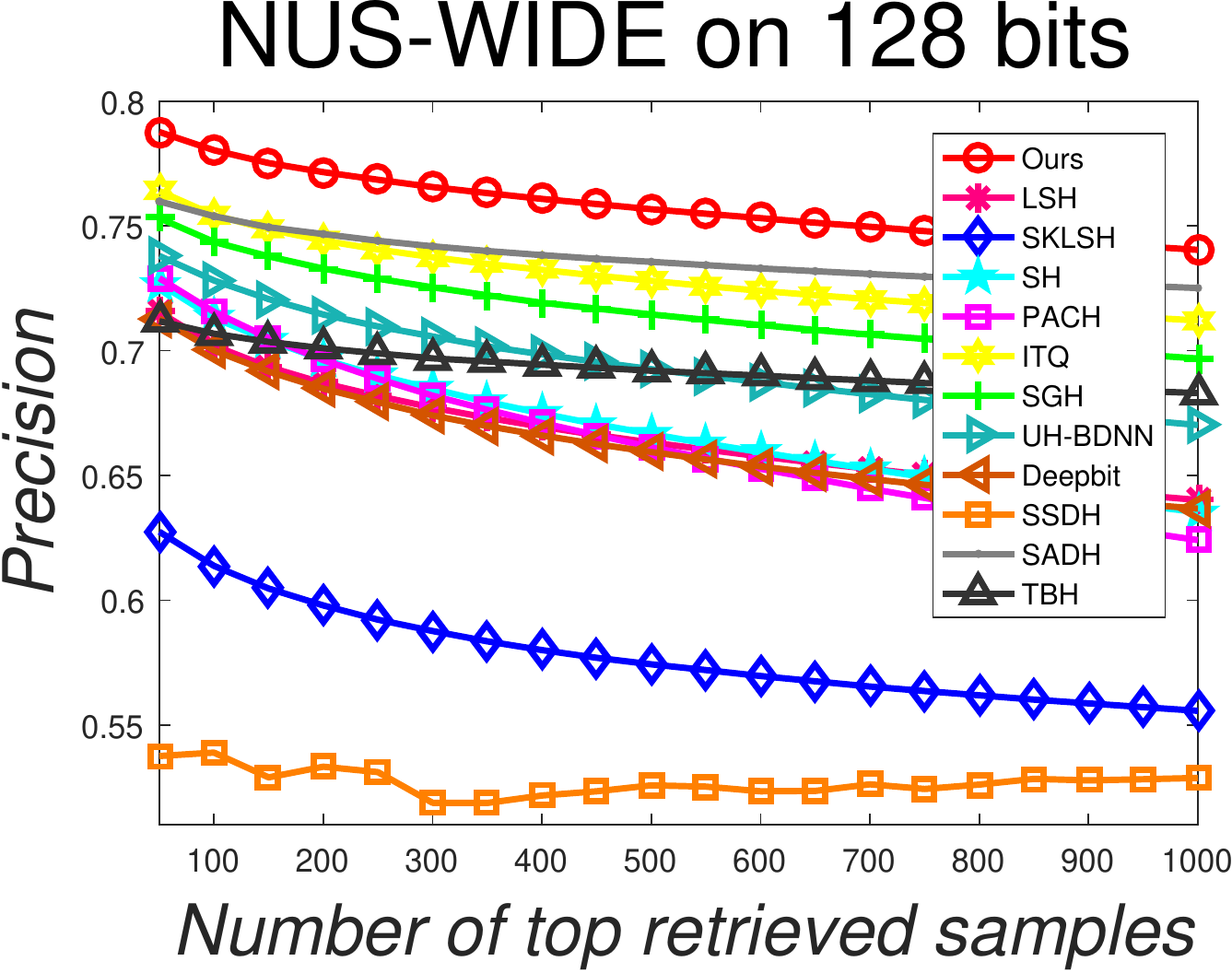}}
\vspace{-4mm}
\caption{The topK-Precision curves on MS COCO and NUS-WIDE.}
\vspace{-3mm}
\label{pre}
%\vspace{-4mm}
\end{figure*}
\vspace{-2mm}
\subsection{Baseline Methods}
We compare our method with several representative unsupervised hashing methods, including
LSH \cite{LSH1999}, SKLSH \cite{SKLSH2009}, SH \cite{SH2008},  % AGH \cite{AGH2011},
PCAH \cite{PCAH2012}, ITQ \cite{ITQ2013}, SGH \cite{SGH2015}, % LSMH \cite{LSMH2017},
UH-BDNN \cite{UH-BDNN2016}, DeepBit \cite{DeepBit2016},
SSDH \cite{SSDH2018}, SADH \cite{SADH2018} and TBH \cite{TBH2020}. % and SCADH \cite{SCADH2020}.
The former six are shallow hashing methods while the latter five and ours are deep hashing methods.
For fair comparisons, the shallow hashing methods and UH-BDNN abandon hand-crafted features and also use deep features extracted by VGG16 model for the experiments.
The source codes of all the baseline methods are provided by their authors.
We carefully tune the parameters of the models to fit our datasets and report their best results for comparison.
\vspace{-2mm}
\subsection{Retrieval Accuracy Comparison}
We compare our method with all the baseline methods to evaluate the retrieval accuracy. The MAP@1000 values and topK-Precision curves with the hash code length varying from 16 to 128 on MS COCO and NUS-WIDE are presented in Table \ref{mapresults} and Figure \ref{pre}, respectively. Based on these results, we have the following observations:

1) The MAP values of our LAGNH method consistently outperform those of the baseline methods on all hash code lengths of both datasets. Particularly on MS COCO dataset, our LAGNH has obvious advantages. The MAP values of different hash code lengths in our method are 0.8476, 0.8655, 0.8718, 0.8751, while the best results in the baselines are 0.6942, 0.7605, 0.7981, 0.8161, respectively. Our method improves the retrieval performance by more than 5\% - 14\% compared with the second best results.
On NUS-WIDE dataset, our method also has 2\% - 3\% improvement of retrieval performance compared with the second best results on four hash code lengths.
The results in Table \ref{mapresults} clearly illustrate the effectiveness of our LAGNH on image retrieval. The superior performance of the proposed method is attributed to the reason that our approach significantly reduces the number of parameters to be optimized and exploits the auxiliary semantics to enhance the representation capability of deep networks.

2) We find that promising MAP results can be obtained by shallow methods using deep features as the model input. The results show that they achieve comparable or better performance than the deep methods. For example, on NUS-WIDE, ITQ with deep features obtains the MAP of 0.6705, 0.6976, and 0.7218 when the hash code length is 16 bits, 32 bits,  and 64 bits respectively, which are the second best results. This phenomenon indicates that the deep neural network has powerful representation capability which can be easily transferred to other models. Also, these results demonstrate that the existing unsupervised deep hashing methods have not exerted the powerful representation capability of the deep model yet. This might be due to the fact that the large amount of parameters in deep neural network models cannot be optimized well without proper semantic supervision under the existing unsupervised deep hash learning framework.

3) From the topK-Precision curves in Figure \ref{pre}, it can be obviously observed that with the increasing number of retrieved images, the retrieval accuracy of our LAGNH consistently outperforms other baselines on both datasets. These figures demonstrate similar trends as the MAP results in Table \ref{mapresults}.

4) We note that Deepbit and SSDH do not perform satisfactorily in our experiments. The reason for DeepBit is that it only enforces three objectives (rotation invariant, quantization loss minimization, and evenly distribution) on the hash layer of the network, without considering the similarity preserving. As for SSDH, the potential reason is that the assumption that the similarity distribution of the sample points is two half-Gaussian distributions does not apply to all datasets.
\vspace{-2mm}
\begin{table}[]
%\footnotesize
\caption{Training and encoding time on MS COCO and NUS-WIDE with 32 bits. The results are reported in seconds.}
\vspace{-3mm}
\begin{tabular}{lcc}
\toprule
Methods & Training time (s) & Encoding time (s) \\ \midrule
\multicolumn{3}{c}{MS COCO}                \\ \midrule
UH-BDNN \cite{UH-BDNN2016}& 6301.7021       & 3.9999       \\
Deepbit \cite{DeepBit2016}& 1802.0045       & 1509.9155    \\
SSDH    \cite{SSDH2018}& 1263.5262       & 4618.4699    \\
SADH    \cite{SADH2018}& 3253.1953       &1478.9802     \\
TBH     \cite{TBH2020}& 4938.3519       & 3.2057       \\
Ours    & 466.3686        & 9.8646       \\ \midrule
\multicolumn{3}{c}{NUS-WIDE}           \\ \midrule
UH-BDNN \cite{UH-BDNN2016} & 6324.5006       & 4.7136       \\
Deepbit \cite{DeepBit2016} & 1858.2623       & 1801.1218    \\
SSDH    \cite{SSDH2018}    & 877.389         & 2630.8497    \\
SADH    \cite{SADH2018}    & 3453.3845       & 1778.8724    \\
TBH     \cite{TBH2020}     & 4945.0841       & 3.266        \\
Ours     & 586.2117        & 11.6804      \\ \bottomrule
\end{tabular}\label{timeresults}
\vspace{-4mm}
\end{table}
\subsection{Run Time Comparison} \label{time}
To demonstrate the advantage of our lightweight network architecture, we conduct experiments to compare the training and encoding time between our LAGNH and the deep baseline methods. Table \ref{timeresults} presents the comparison results on MS COCO and NUS-WIDE when the code length is fixed to 32 bits. From the experimental results, we can observe that the advantages of our method on training efficiency are obvious. The training time of LAGNH is 466.3686 seconds on MS COCO, and 586.2117 seconds on NUS-WIDE. They are at least 8.4x faster than UH-BDNN and TBH, and 5.8x faster than SADH. The encoding time of our method is similar to those of UH-BDNN and TBH, and is much faster than Deepbit, SSDH and SADH.
The improved efficiency of our LAGNH is mainly because that, we design the lightweight attention module and adversarial regularized graph convolutional network with the assistance of auxiliary semantics, which greatly reduces the network parameters to be optimized at the training phase and the nonlinear transformations to be carried out at the testing phase.

\begin{figure}
\centering
\subfigure{\includegraphics[scale=0.29]{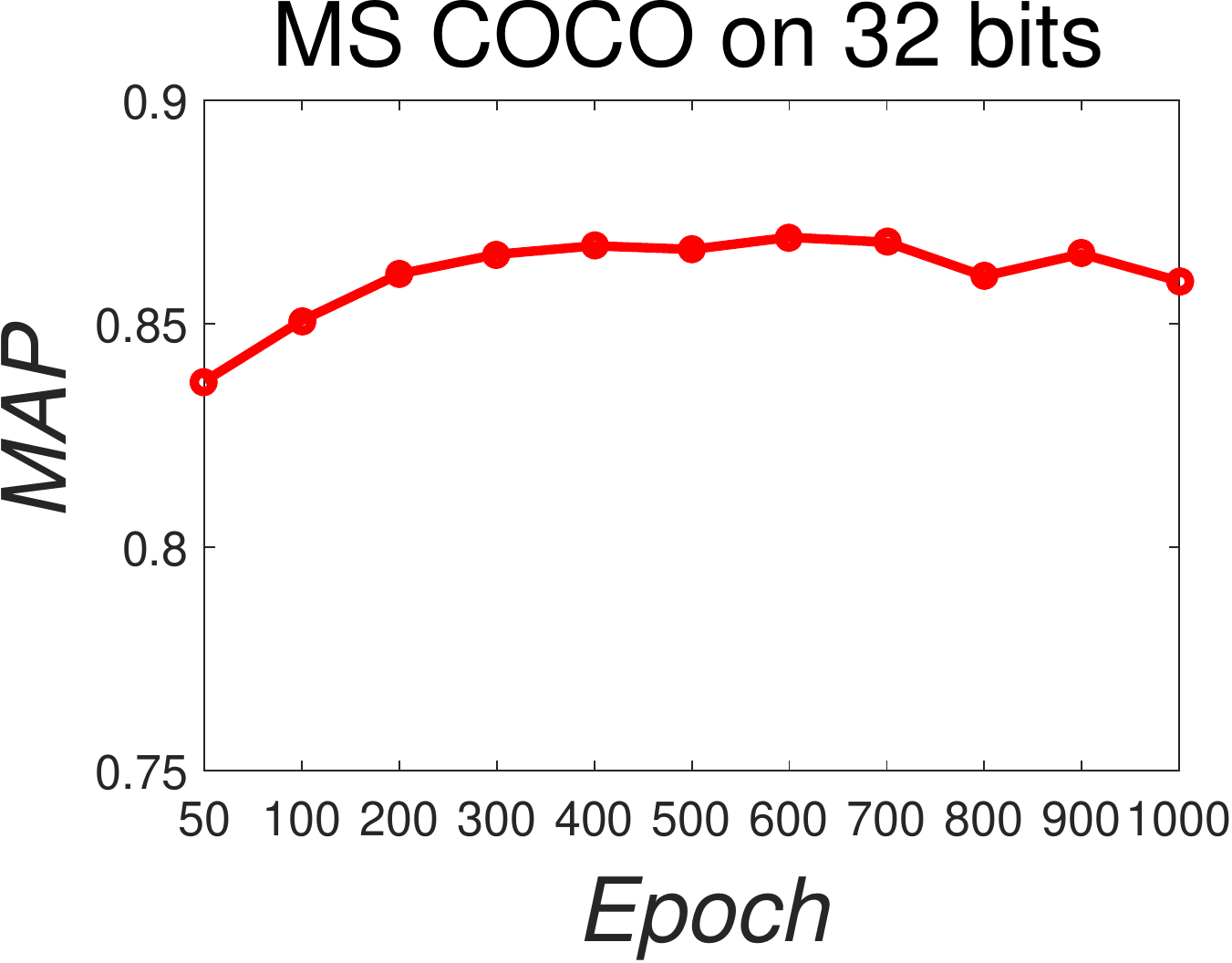}}
\subfigure{\includegraphics[scale=0.29]{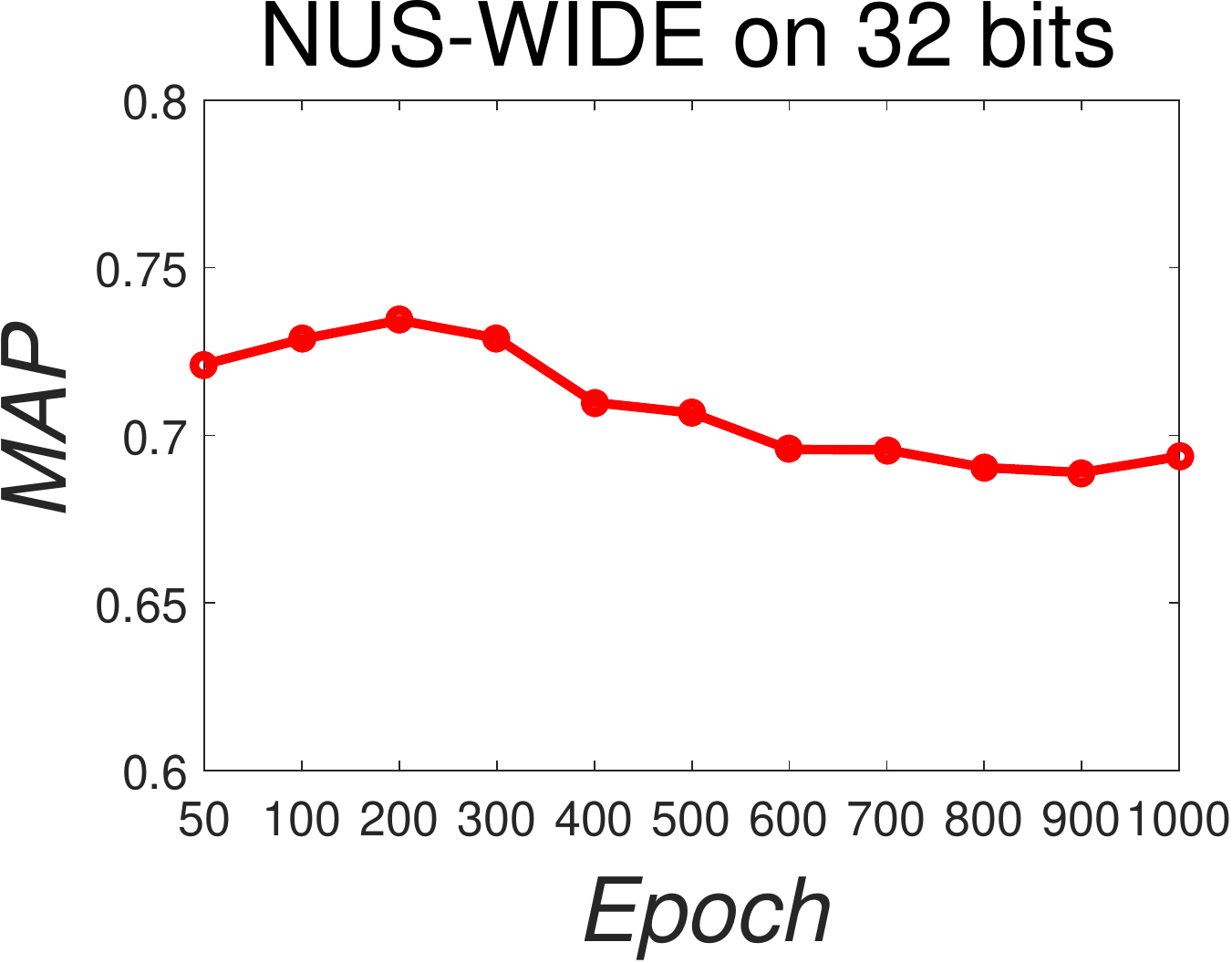}}
\vspace{-3mm}
\caption{Variations of MAP value with the number of training epochs.}
\vspace{-3mm}
\label{epochmap}
\end{figure}

In addition, we conduct experiments to verify the impact of the training epochs on the performance of LAGNH.
We report the MAP values for different epochs on MS COCO and NUSWIDE when the code length is fixed to 32 bits. The number of epochs starts at 50 and gradually increases from 100 to 1000 with a step size of 100. From Figure \ref{epochmap}, we can see that our method can achieve satisfactory performance with fewer epochs (about 300 epochs) on both datasets. These experimental results show that our LAGNH method can increase the MAP value quickly, and thus has the advantage on training efficiency.
\begin{table*}[]
%\footnotesize
\caption{MAP comparison results of the variants of our approach on MS COCO and NUS-WIDE.}
\vspace{-3mm}
\begin{tabular}{lllllllll}
\toprule
\multirow{2}{*}{Methods} & \multicolumn{4}{c}{MS COCO}           & \multicolumn{4}{c}{NUS-WIDE} \\ \cline{2-9}
                         & 16 bits & 32 bits & 64 bits & 128 bits & 16 bits & 32 bits & 64 bits & 128 bits \\ \hline\hline
LAGNH-no-aux	&	0.6163 	&	0.5704 	&	0.5550 	&	0.5726 	&	0.5302 	&	0.4979 	&	0.3780 	&	0.5435 	\\		
LAGNH-no-att	&	0.8424 	&	0.8631 	&	0.8681 	&	0.8746 	&	0.7017 	&	0.7190 	&	0.7490 	&	0.7581 	\\		
LAGNH-only-S$^{v}$	&	0.7279 	&	0.7578 	&	0.7539 	&	0.7672 	&	0.5774 	&	0.5677 	&	0.5899 	&	0.6193 	\\		
LAGNH-only-S$^{a}$	&	0.7961 	&	0.8121 	&	0.8192 	&	0.8156 	&	0.5981 	&	0.5962 	&	0.6075 	&	0.6068 	\\	\hline	
LAGNH-recons-Z$^\texttt{T}$Z	&	0.7927 	&	0.8210 	&	0.8407 	&	0.8514 	&	0.5443 	&	0.5192 	&	0.5956 	&	0.5850 	\\		
LAGNH-recons-feat	&	0.7690 	&	0.7955 	&	0.8232 	&	0.8399 	&	0.6747 	&	0.6872 	&	0.7374 	&	0.7501 	\\	\hline
LAGNH-recons-S	&	0.7312 	&	0.7317 	&	0.7329 	&	0.7586 	&	0.7011 	&	0.7226 	&	0.7503 	&	0.7560 	\\		
LAGNH-recons-S$^{v}$	&	0.7402 	&	0.7079 	&	0.7019 	&	0.7080 	&	0.6946 	&	0.7303 	&	0.7515 	&	0.7539 	\\		
LAGNH-recons-S$^{a}$(Ours) 	&	0.8476 	&	0.8655 	&	0.8718 	&	0.8751 	&	0.7061 	&	0.7242 	&	0.7530 	&	0.7644 	\\	
\bottomrule
\end{tabular}\label{ablaresults}
\vspace{-4mm}
\end{table*}
\subsection{Model Analysis}
\subsubsection{Discussion of different model components}
To verify the effectiveness of different components in our model, we design several variants for comparison:
\begin{itemize}
  \item \emph{LAGNH-no-aux}: removes the auxiliary semantic information in the whole framework.
  \item \emph{LAGNH-no-att}: removes the attention component and uses deep image features and auxiliary semantic information directly to construct the augmented semantic graph.
  \item \emph{LAGNH-only-S$^{v}$}: only feeds the visual similarity graph to the GCN.
  \item \emph{LAGNH-only-S$^{a}$}: only feeds the auxiliary similarity graph to the GCN.
\end{itemize}

The comparative results on MAP are shown in Table \ref{ablaresults}.
From the results, we find that our method performs better than the other four variants on two datasets with the code length varying from 16 bits to 128 bits.
Particularly, the largest MAP drop is caused when no auxiliary semantic information is exploited for guiding the network.
Using only the visual similarity graph or the auxiliary similarity graph can also result in obvious performance degradation. These results validate that our architecture indeed strengthens the important information and captures the complicated relationships of images, which can help our model to learn more robust hash codes.
\vspace{-1mm}
\subsubsection{Discussion of similarity graph reconstruction}
For reconstruction, we aim to discuss the following three questions:
\begin{description}
\item[Q1:] How about the performance of the proposed method if we directly reconstruct the augmented semantic graph by the outputs of the GCN?
\item[Q2:] How about the performance of the proposed method if we reconstruct the original images?
\item[Q3:] How about the performance of the proposed method if different similarity graphs are reconstructed?
\end{description}

\vspace{-1mm}
We design the variant approaches \emph{LAGNH-recons-Z$^\texttt{T}$Z} and \emph{LAGNH-recons-feat} to answer \textbf{Q1} and \textbf{Q2}, respectively.
\emph{LAGNH-recons-Z$^\texttt{T}$Z} minimizes the augmented semantic graph and the inner products of the outputs of the GCN, it can be formulated as
\begin{equation} \nonumber
\min \mathcal L_{recons} = {\left\| {k{\mathbf{S}^{a}} -  {{\mathbf{Z}}^{\texttt{T}}}{\mathbf{Z}}} \right\|_\texttt{F}^2}.
\end{equation}
\emph{LAGNH-recons-feat} imports the outputs of the GCN into a decoder, which generates reconstructed image features.
We minimize the errors between the original image features and the reconstructed image features, that is
\begin{equation} \nonumber
\min \mathcal L_{recons} = {\left\| { \mathbf{X} - \hat{ \mathbf{X}}} \right\|_\texttt{F}^2}.
\end{equation}

The experiments are conducted on MS COCO and NUS-WIDE with the hash code length varied from 16 bits to 128 bits.
Table \ref{ablaresults} presents the MAP comparison results.
The results show that neither the reconstruction of the augmented semantic graph by the inner products of the outputs of the GCN nor the direct reconstruction of the original image features can achieve satisfactory performance.

In addition, to verify which graph is the best choice for reconstruction, we test the following schemes:
\begin{itemize}
\item \emph{LAGNH-recons-S}: reconstructs the augmented semantic graph.
\item \emph{LAGNH-recons-S$^{v}$}: reconstructs the visual similarity graph.
\item \emph{LAGNH-recons-S$^{a}$}: reconstructs the auxiliary similarity graph.
\end{itemize}
\vspace{-1mm}
The experimental results are also listed in Table \ref{ablaresults}.
We can see from the results, on NUS-WIDE, the impacts of the reconstruction of different graph structures on performance are relatively small, while on MS COCO, the performance changes are large.
This may be because the auxiliary semantic information of the two datasets comes from different sources.
The semantics of the detected objects from MS COCO are more accurate than those in the tags from NUW-WIDE.
In order to unify the framework, we optimize our model by reconstructing the auxiliary similarity graph, because it can obtain relatively good results on both datasets.
\begin{figure}
\centering
\subfigure{\includegraphics[scale=0.2]{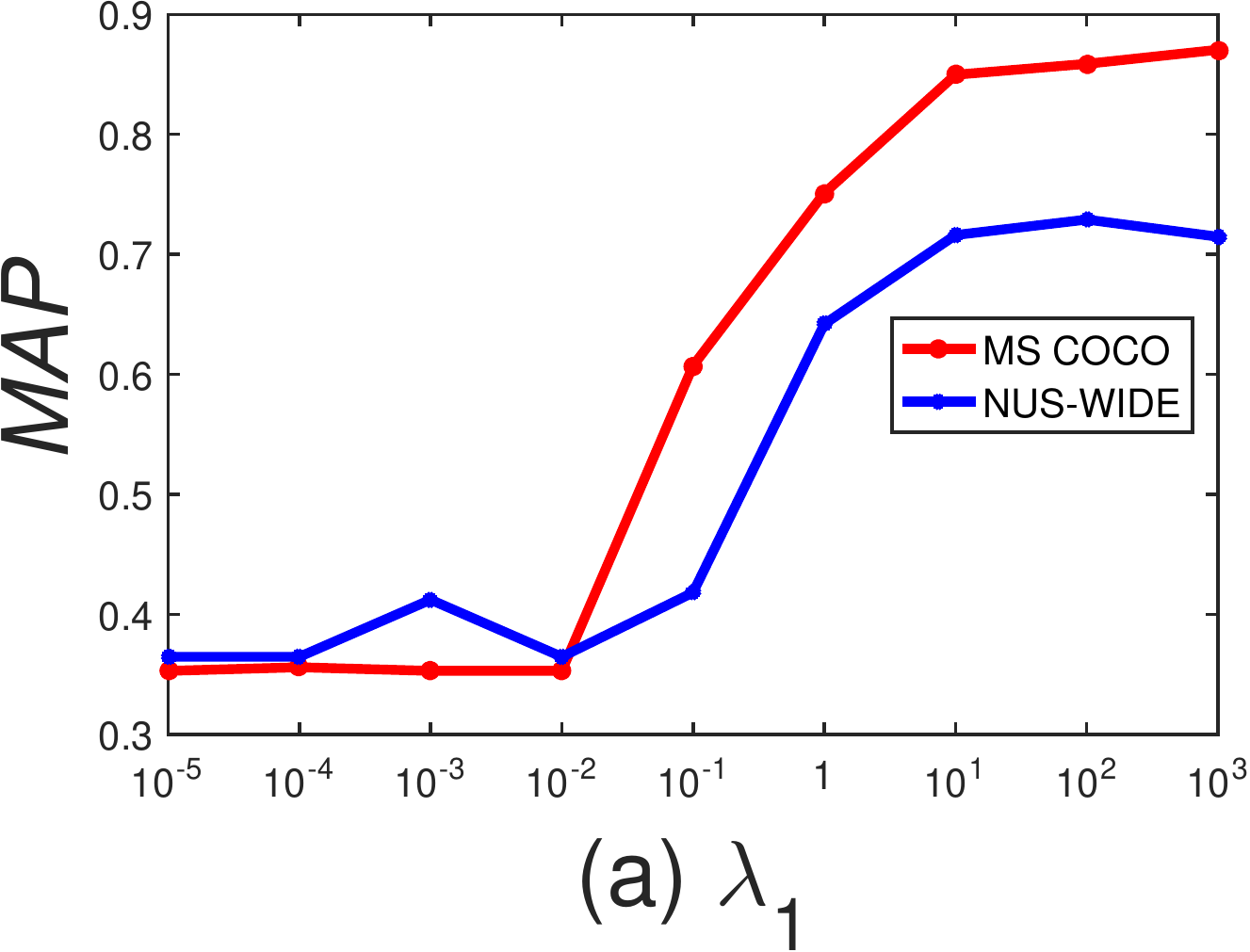}}
\subfigure{\includegraphics[scale=0.2]{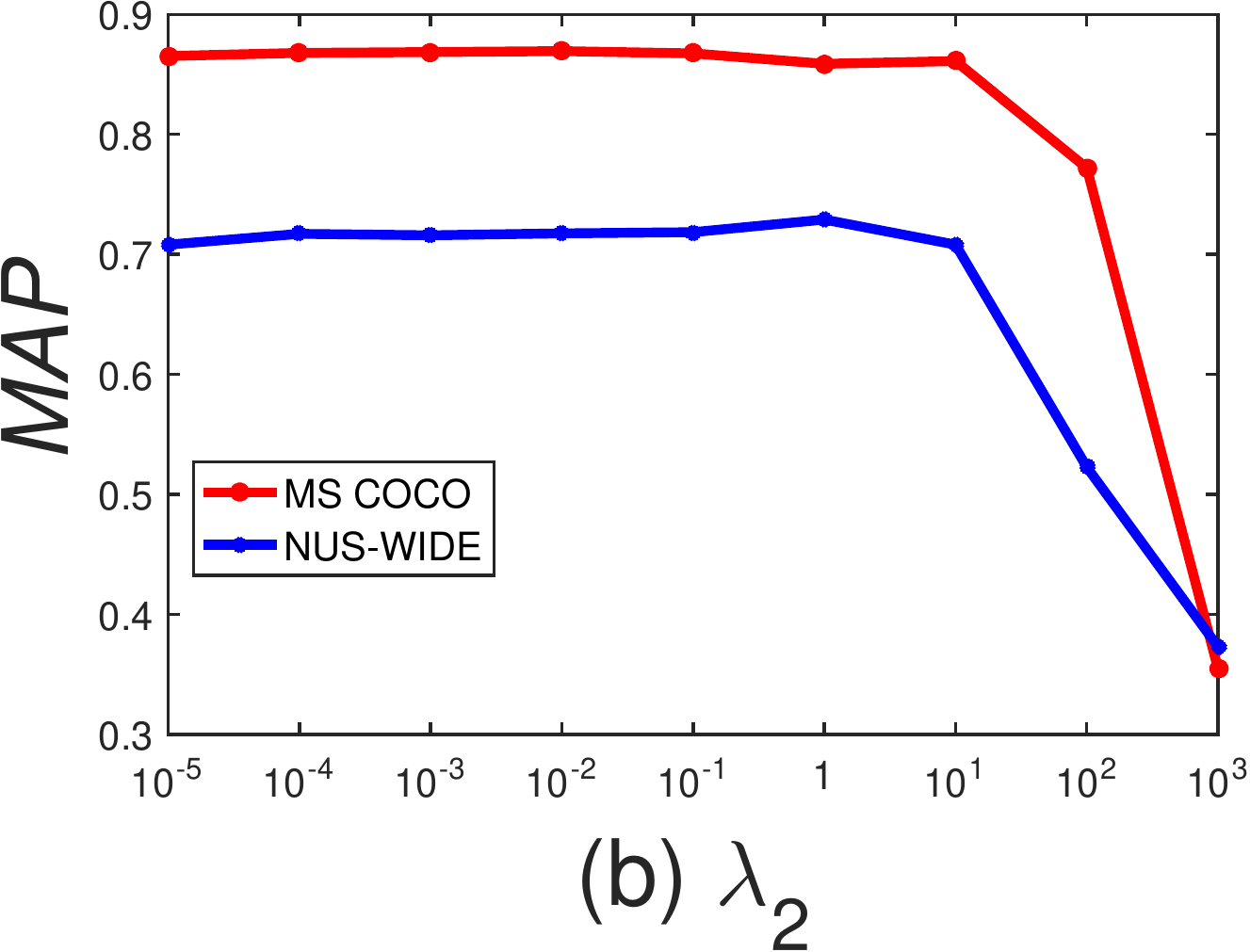}}
\subfigure{\includegraphics[scale=0.2]{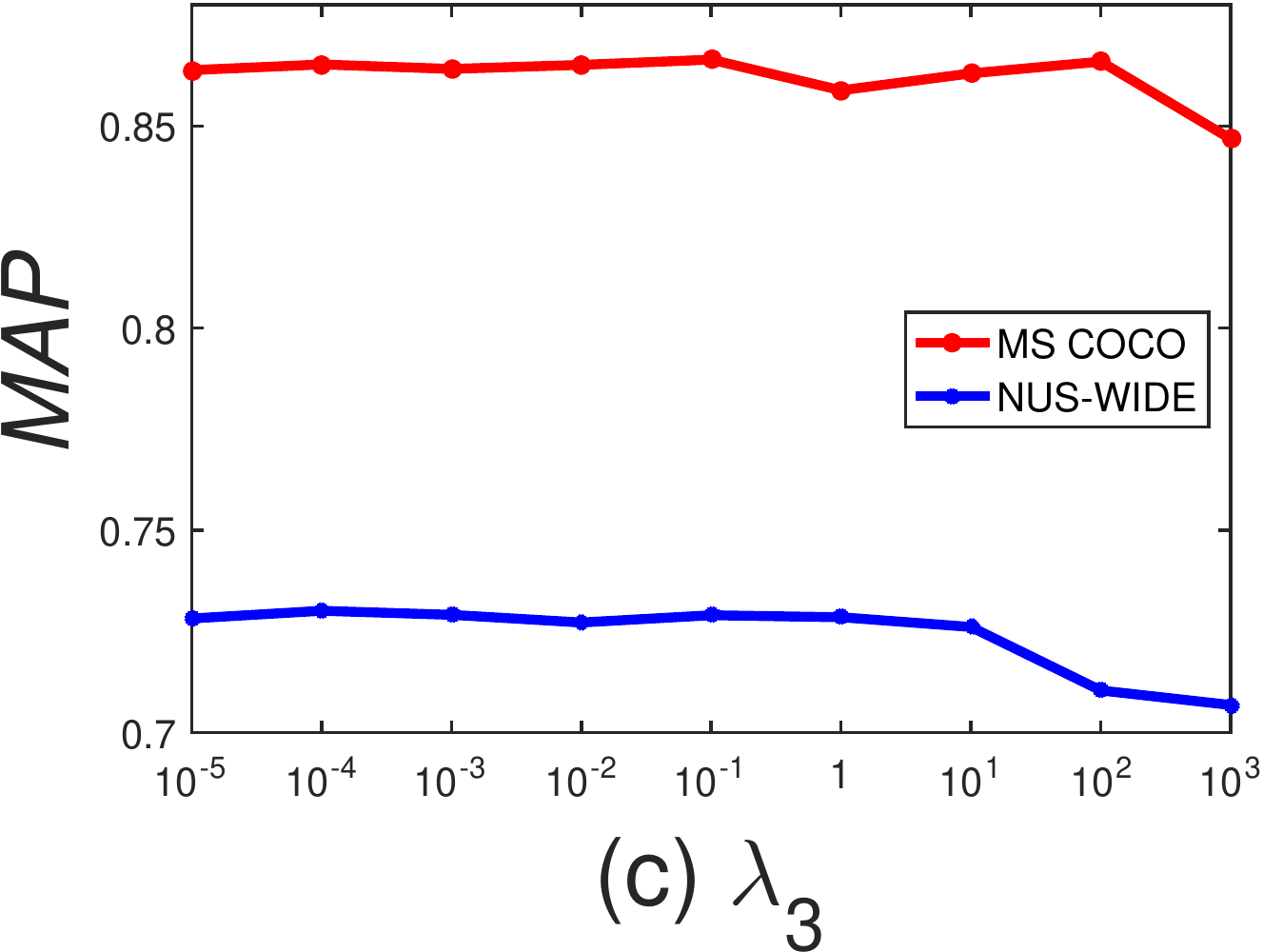}}
\vspace{-5mm}
\caption{MAP variations with respect to different parameters on MS COCO and NUS-WIDE on 32 bits.}
\vspace{-3mm}
\label{para}
\end{figure}
\subsection{Parameter Analysis}
We conduct a sensitivity analysis of the parameters $\lambda_1$, $\lambda_2$, $\lambda_3$ on MS COCO and NUS-WIDE when the hash code length is fixed as 32 bits. Similar results can be found with other code lengths.
The curves in Figure \ref{para} show that our method obtains relatively stable retrieval performance when
$\lambda_1$ is in the range of $\{10^{1}$, $10^{3}\}$, $\lambda_2$ is in the range of $\{10^{-5}$, $1\}$, $\lambda_3$ is in the range of $\{10^{-5}$, $10^{2}\}$ on MS COCO,
and $\lambda_1$ is in the range of $\{10^{1}$, $10^{3}\}$, $\lambda_2$ is in the range of $\{10^{-5}$, $1\}$, $\lambda_3$ is in the range of $\{10^{-5}$, $10^{1}\}$ on NUS-WIDE, respectively.
\section{Conclusion}
In this paper, we propose a simple yet effective unsupervised deep hashing method, termed \emph{Lightweight Augmented Graph Network Hashing} (LAGNH) with a two-pronged strategy, to avoid the large amount of parameters in deep hashing networks and the shortage of semantic supervision in unsupervised hashing. We employ the auxiliary semantics freely extracted from the inner structure of the image as the supervision to guide the training process of hash learning, which enhances the semantic representation capability of the deep network. Simultaneously, we design a lightweight network structure with the assistance of the auxiliary semantics, which significantly reduces the number of parameters to be optimized and thus greatly speedups the training process. Experiments on two public image retrieval datasets demonstrate the superior performance of our proposed two-pronged strategy.
\begin{acks}
This work is supported in part by the National Natural Science Foundation of China (Nos. 61802236, U1836216),
in part by the Natural Science Foundation of Shandong, China (Nos. ZR2020YQ47, ZR2019QF002, ZR2019ZD03),
and in part by the Youth Innovation Project of Shandong Universities, China (No. 2019KJN040).
\end{acks}

%%
%% The next two lines define the bibliography style to be used, and
%% the bibliography file.
\bibliographystyle{ACM-Reference-Format}
\bibliography{sample-base}

%%
%% If your work has an appendix, this is the place to put it.
%\appendix

\end{document}